\documentclass[english,article]{revtex4}
\usepackage[caption=false]{subfig}
\usepackage[latin9]{inputenc}
\usepackage{amsmath}
\usepackage{graphicx}
\usepackage{amssymb}
\newcommand{\ket}[1]{| #1 \rangle}
\newcommand{\bra}[1]{\langle #1 |}

\begin{document}

\title{Nonlinear spectroscopy of superconducting anharmonic resonators}

\author{D.P. DiVincenzo}

\email{d.divincenzo@fz-juelich.de} 
\affiliation{RWTH Aachen and Forschungszentrum Juelich, Germany}
\author{J.A. Smolin}
\email{smolin@watson.ibm.com}
\affiliation{IBM T.J. Watson Research Center, Yorktown Heights, NY 10598 USA}


%
\begin{abstract}

  We formulate a model for the steady state response of a nonlinear
  quantum oscillator structure, such as those used in a variety of
  superconducting qubit experiments, when excited by a steady, but not
  necessarily small, ac tone.  We show that this model can be derived
  directly from a circuit description of some recent qubit experiments
  in which the state of the qubit is read out directly, without a
  SQUID magnetometer.  The excitation profile has a rich structure
  depending on the detuning of the tone from the small-signal resonant
  frequency, on the degree of damping, and on the excitation
  amplitude.  We explore two regions in detail: First, at high damping
  there is a trough in the excitation response as a function of
  detuning, near where the classical Duffing bifurcation occurs.  This
  trough has been understood as a classical interference between two
  metastable responses with opposite phase.  We use Wigner function
  studies to show that while this picture is roughly correct, there
  are also more quantum mechanical aspects to this feature.  Second,
  at low damping we study the emergence of sharp, discrete spectral
  features from a continuum response.  We show that these the
  structures, associated with discrete transitions between different
  excited-state eigenstates of the oscillator, provide an interesting
  example of a quantum Fano resonance.  The trough in the Fano
  response evolves continuously from the ``classical" trough at high
  damping.

\end{abstract}

\maketitle









\section{introduction}

Spectroscopic studies of superconducting qubit systems have increased
steadily in sophistication in recent years.  For many of the first
Josephson-junction qubits, their only link with the outside world was
magnetometry: diamagnetic currents, particularly in flux qubits, were
detectable by SQUID magnetometers.  While this provided the basis for
some of the first experiments in which solid state qubit states were
detected, these magnetometers were very intrusive devices in these
systems, and were clearly an impediment to scalable systems.
Spectroscopy with these SQUID detectors was very limited.

With the application in recent years of many sophisticated techniques
from microwave engineering, various alternatives to the SQUID approach
have emerged.  The detection, protection, and manipulation of
superconducting qubits have undergone impressive improvements.  In
particular, the spectroscopic probing of qubits and qubit/resonator
systems is now capable of very useful high-fidelity qubit
detection\cite{SQUIDless,TroughExpt}, as well as detailed
characterization of the parameters of these systems opening
a beautiful view \cite{TroughTheory} into the rich quantum response that
these systems exhibit.

This paper will provide a theoretical study of a model that is central
to these spectroscopic studies.  The model is that of a single
oscillator (bosonic mode) with a purely quartic anharmonicity.  The
oscillator is driven by a single ac tone, which is not necessarily a
small perturbation on the system, and the system is damped with a
purely frictional term, representative of a zero-temperature quantum
environment.  The observable is taken to be essentially the
displacement of this resonator.

We will emphasize here the insights that can be gained from analytic
or semi-analytic studies of this model.  In previous studies of this
model\cite{DW}, it has been shown that the steady state response of
the system can be calculated in closed form.  It is very rewarding to
look at the many aspects of this amazingly rich solution.  At high
damping it exhibits much of the phenomenology of the classical Duffing
oscillator.  In a broad region of detuning, on the ``red" side of the
harmonic response frequency, and for moderately high amplitude
excitation, the response is ``high", or quasi-resonant.  But as
damping is decreased, new, sharp resonant structures emerge from the
broad Duffing feature.  These lines are clearly associated with
individual spectroscopic transitions in the anharmonic ladder of
energy eigenstates.  But the line shapes of these new lines are very
non-Lorentzian, and are indicative of a complex interaction among
different response channels.

In the present work we do further calculations to gain more
understanding of several of these spectroscopic features.  Up to a
point, the behavior in the high damping regime exactly matches the
classical Duffing theory.  There is a linear response regime, in which
the response is a simple Lorentzian centered on the harmonic frequency
of the oscillator; at higher excitation amplitude the resonant line
moves to lower frequency and sharpens. But there are differences: the
quantum model intrinsically does not have multiple solutions for the
steady state response, and so does not have any hysteretic
behavior. Near the classical bifurcation point the quantum response
shows a fairly abrupt transition, as a function of excitation
detuning, between a ``low" and ``high" response regime.

While some details of this interesting quantum behavior have been
discussed in previous work, we shed new light on this behavior using
simple Wigner function computations.  In the middle of this transition
from ``low" to ``high" response, there is a dip in the response
amplitude, which has been interpreted \cite{DW,TroughTheory} as a kind
of intermittency in the steady state involving two responses with
different, and therefore canceling, phases.  Our Wigner functions show
that there is clearly some truth to this point of view, but we find
that, even for damping large enough to wash out all other quantum
spectral features, the dip has more features than can be explained by
any classical intermittency point of view.  We find that this story is
considerably enriched by using some of the modern tools of quantum
information theory, such as calculations of the von-Neumann entropy,
and some basic facts about the eigenanalysis of the completely
positive map describing time evolution in our model.

We also explore in detail how structure emerges as the damping
parameter is decreased.  While the spectral response grows
considerably in complexity, we find that the simple ``dip" described
above in the high damping regime persists for all values of damping,
but comes in a different guise in this quantum limit.  In one part of
the spectrum, we identify this dip as a feature of a Fano resonance.
The Fano phenomenon is a quintessentially quantum interference, yet we
find that in this case it is one extreme of a family of interference
phenomena that also include a nearly classical form of interference at
the other extreme.

These studies help us to answer a very fundamental question about anharmonic resonant structures, which is: when is it reasonable to call such a structure a qubit?  It is evident that the answer is a dynamical one: the very same resonator can be either a very well defined two-level quantum system (a qubit), or an oscillator requiring a many-level quantum description, depending on the amplitude and frequency of oscillation.  In this paper we will fairly precisely demarcate the rather complex boundary between these two regimes. 

\section{the electrical device}

\subsection{Circuit basics}

\begin{figure}[htp]
	\centering
		\includegraphics[angle=90,width=12cm]{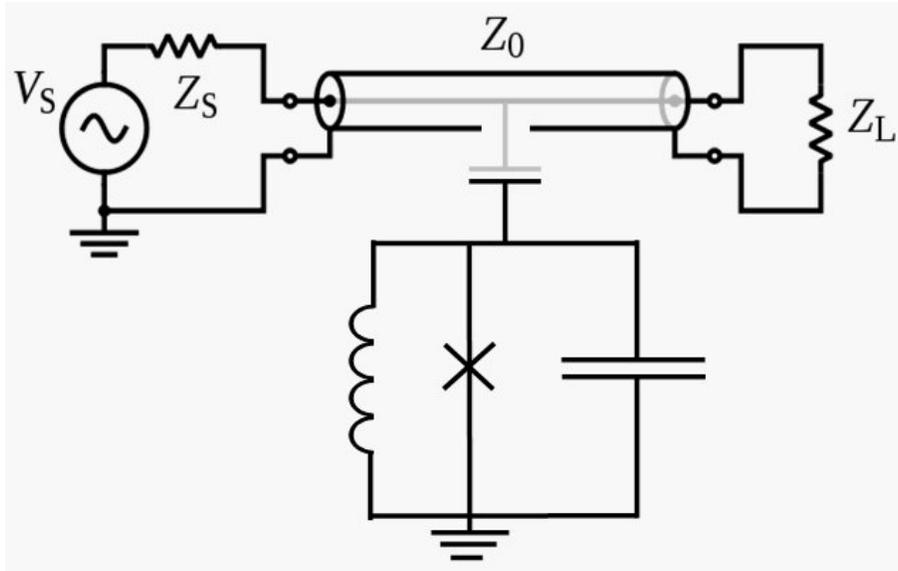}
	\caption{Experimental schematic.}
	\label{fig:fig0}
\end{figure}

The calculations here are motivated by current experiments on
superconducting circuits such as the one shown in Fig. \ref{fig:fig0}.
The ac voltage signal (in the microwave frequency range, in cases of
current interest) excites a transmission line; this transmission line
is coupled capacitively to a small, discrete superconducting circuit
that we will, for short, call ``the qubit."  But an important message
of this paper is that structures of interest do not always behave as
qubits, that is, like two-level quantum mechanical systems.  From the
point of view of electrical science, the combination shown of a
capacitor, inductor, and Josephson junction is viewed as an anharmonic
resonator.  As we will show, depending on the excitation conditions,
this classical view is sometimes more appropriate than the quantum
mechanical one.  In fact, there is a rich set of phenomena that are
explained by neither the classical or the simple (two-level) quantum
point of view.

Many features of a real experiment are omitted from
Fig. \ref{fig:fig0}.  Naturally the qubit is held at low temperature,
while the microwave voltage source is (presently) always at room
temperature; the diagram does not show the attenuator stages that are
necessary to keep room temperature noise out of the qubit.  But these
stages are engineered so that the functioning is that of a cold,
matched ({\em i.e.}, $Z_S=Z_0$) source.  On the output end,
conceptually measurements of the qubit are recorded as the voltage
across the terminating resistor $Z_L$.  In reality, this circuit is a
complex, active chain of amplifiers and other components, with a
signal recorded at room temperature.  But again, the engineering of
this chain has the object of reducing the effective functioning to
that of the simple circuit in Fig.  \ref{fig:fig0}, in which the load
is matched ($Z_L=Z_0$) and cold.

Much experiment and analysis (\emph{cf.} \cite{saturatetoo}) has been
devoted to a related circuit, in which the transmission line is only
weakly (capacitively) coupled to the source and load circuits.  This
is the ``circuit QED" setting, in which the transmission line forms a
resonator which functions as a quantum-coherent resonator.  The
combination of this quantum resonator and the qubit exhibits rich,
interesting, and potentially useful physics.  The system we analyze
here does not have all the same potentialities as the cQED system, but
its dynamics still has a great deal of complexity in it, and the
simpler circuit has its own potential application in the realm of
simple, reliable qubit measuring circuits.

\begin{figure}[htp]
	\centering
		\includegraphics[width=14cm]{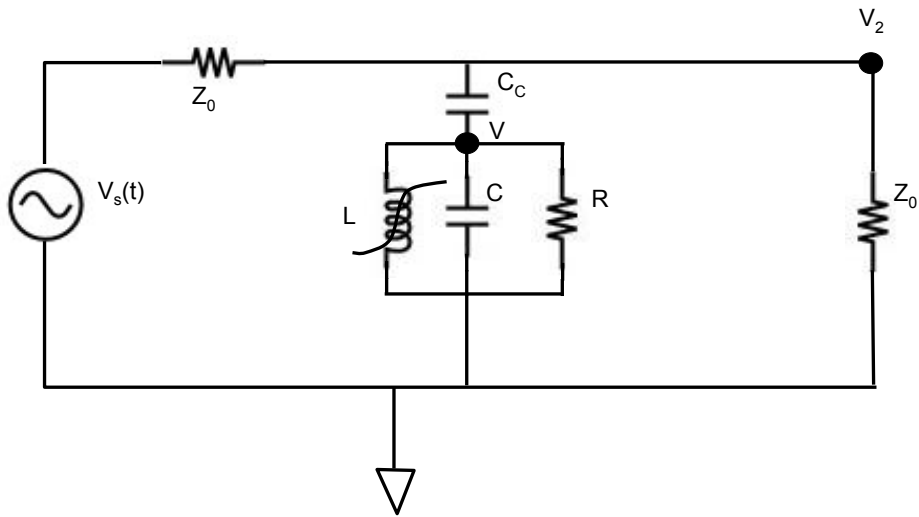}
	\caption{Circuit for experiment.  $Z_0=50\Omega$.}
	\label{fig:fig1}
\end{figure}

As far as the dynamics of the qubit is concerned, the circuit of
Fig. \ref{fig:fig1} accurately describes the schematic of
Fig. \ref{fig:fig0} in the matched condition $Z_0=Z_S=Z_L$.  The
voltage $V_2$ at the node indicated is the observable of the actual
experimental setup.  We have replaced the inductor and Josephson
junction by a nonlinear inductor, which is sufficient given that we
will not wish to describe the global response of the device ({\em
  e.g.}, to large changes of an external bias flux).  We have also
included resistor $R$, to describe internal loss of the qubit.  This
is in some cases competitive with the losses due to the ``external"
resistors in the circuit of Fig. \ref{fig:fig1}.

\begin{figure}[htp]
	\centering
		\includegraphics[width=14cm]{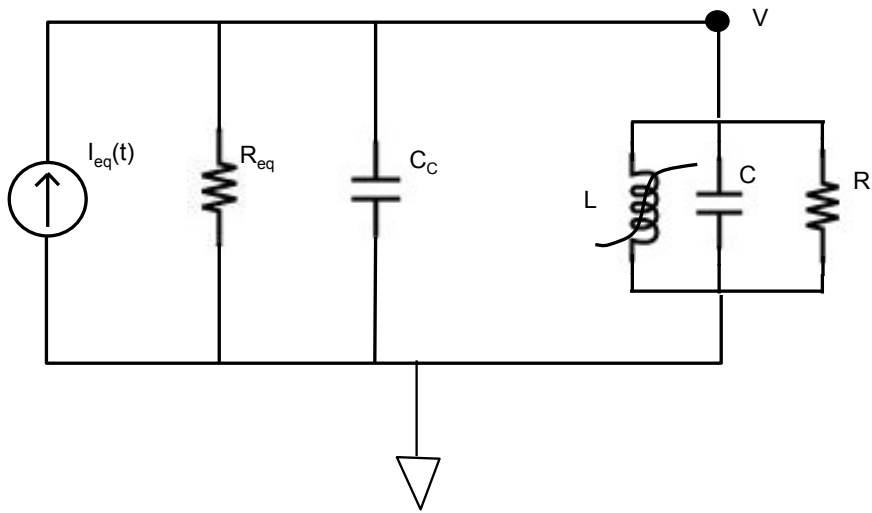}
	\caption{Equivalent circuit to Fig. \ref{fig:fig1}.  Note that $R_{eq}=2Z_0/(\omega_0C_cZ_0)^2$ with $\omega_0=1/\sqrt{LC}$  
and $I_{eq}=V_s/Z_0$.  The voltage $V$ is then identical in the two circuits.}
	\label{fig:narrow}
\end{figure}

We can get another circuit more directly amenable to a quantum
treatment by using some circuit transformations due to Zmuidzinas {\em
  et al.}\cite{Zmunpub}, and making a ``narrow band" approximation,
which amounts to saying that the only response of interest in the
Fourier transform of the voltage will be at frequencies near the
excitation frequency; this will be compatible with the rotating wave
approximation that we will introduce later.  This is shown in
Fig. \ref{fig:narrow}.

Note that the measurable quantity $V_2$ has disappeared from this circuit.  But is easily recovered from the voltage $V$ via the linear circuit relation\cite{Zmunpub}
\begin{equation}
V_2={1\over 2}V_s+(i\omega_0C_cZ_0)V\label{lrelate}
\end{equation}
This is valid for $\omega_0C_cZ_0<<1$.  Eq. (\ref{lrelate}) is understood to be in the frequency domain, with the ``{\em i}" indicating that the contribution of $V$ to the signal $V_2$ is out of phase with that of the drive $V_s$.

We will take $V_s$ to be a classical ac variable with a simple sinusoidal time dependence:
\begin{equation}
V_s(t)=V_s\cos(\omega_pt)
\end{equation}
We will assume that the nonlinear inductance is characterized by the small-signal energy storage formula
\begin{equation}
E_L={1\over 2}{\Phi^2\over L}+{1\over 4}{\Phi^4\over L_4}
\end{equation}
$\Phi$ is the (formal) magnetic flux through this inductor, {\em i.e.}, $\Phi(t)=\int^tV_L(t')dt'$.  Other powers of $\Phi$ will generally appear in this energy, but our truncation will often be very accurate.  This energy expression implies the non-linear inductive two-terminal response
\begin{equation}
I_L={\partial E_L\over\partial\Phi}={\Phi\over L}+{\Phi^3\over L_4}\ .
\end{equation}
Using any standard circuit mechanics treatment ({\em e.g.}, \cite{BKD}), the Hamiltonian of the circuit of Fig. \ref{fig:narrow} is then
\begin{equation}
H={1\over 2}{Q^2\over C'}+\left({1\over 2}{\Phi^2\over L}+{1\over 4}{\Phi^4\over L_4}\right)+{V_s(t)\over Z_0}\Phi
\end{equation}
Here $C'=C+C_c$.  The classical equation of motion for $\Phi$ will also have a ``friction" term going like 
\begin{equation}
-{{\dot\Phi}\over R'},\,\,\,\,\,\,{1\over R'}={1\over R_{eq}}+{1\over R},\,\,\,\,\,R_{eq}={2Z_0\over(\omega_0C_cZ_0)^2}
\end{equation}

Note that the voltage $V$ of interest here is related to our canonical momentum via the two-terminal relation for the linear capacitive element:
\begin{equation}
V={Q\over C'}\ .
\end{equation}

\subsection{Quantum treatment}
We go over now to a quantum treatment of the circuit.  This will give
$Q$, and therefore $V$ and $V_2$, the status of a quantum operators.
We will assume that the experiment consists of a sequence of weak
measurements of $V_2$, so that we will extract the expectation value
$\langle V_2\rangle$, and that the system is not otherwise disturbed
by the measurement.  The standard quantum commutation relations are
\begin{equation}
[\Phi,Q]=i\hbar
\end{equation}
Introducing bosonic creation and annihilation operators, we have
\begin{equation}
Q=-i\sqrt{\hbar\over 2}\left({C'\over L}\right)^{1\over 4}(a_{\rm Sch}-a_{\rm Sch}^\dagger),\label{theQ}
\end{equation}
and similarly for $\Phi$. Here the subscript on the operator $a$ stands for ``Schrodinger picture,'' as we will soon do most of our calculations in an interaction picture.

Using this operator expression for $Q$ in Eqs. (\ref{theQ}), we can
get a quantum expression for the observable $V_2$.  Going to the time
domain:
\begin{equation}
\langle V_2\rangle={1\over 2}V_s\cos\omega_p t+{\omega_0C_cZ_0\over C'}({\rm Re}\langle a\rangle\cos\omega_pt+{\rm Im}\langle a\rangle\sin\omega_pt)\label{lrelate2}
\end{equation}
The Hamiltonian operator describing the coherent part of the system
evolution becomes
\begin{equation}
H_S=\hbar\omega_0a_{\rm Sch}^\dagger a_{\rm Sch}+{\hbar\chi\over 6}(a_{\rm Sch}+a_{\rm Sch}^\dagger)^4
+2\hbar\epsilon\cos(\omega_p t)(a_{\rm Sch}+a_{\rm Sch}^\dagger)
\end{equation}
We now go to an interaction picture ($a=e^{-i\omega_p t}a_{\rm Sch}$, {\em etc.}), and perform a rotating-wave approximation, obtaining
\footnote{We note the useful identity $(a^\dagger)^2a^2\ket{n}=n(n-1)\ket{n}$.}
\begin{equation}
H=\hbar\Delta a^\dagger a+\hbar\chi (a^\dagger)^2a^2
+\hbar\epsilon(a+a^\dagger)\label{RWAeq}
\end{equation}
Here the detuning $\Delta$ is $\Delta=\omega_0-\omega_p$, and we introduce scaled parameters, all with units sec$^{-1}$:
\begin{equation}
\omega_0={1\over\sqrt{LC'}},\,\,\,\chi={3\hbar\over 8}{L\over C'L_4},\,\,\,\,
\epsilon={1\over\sqrt{8\hbar}}\left({L\over C'}\right)^{1\over 4}{V_s\over Z_0}.
\end{equation}
The lossy parts of our circuit are accounted for using the usual bosonic bath description; making the standard Born-Markov approximations \cite{Leggett,BKD}, and assuming the bath to be at zero temperature, the system dynamics is described by the Lindblad master equation:
\begin{equation}
{d\rho\over dt}=-{i\over\hbar}[H,\rho]+{\gamma\over 2}(2a\rho a^\dagger-a^\dagger a\rho-\rho a^\dagger a)
\end{equation}
with
\begin{equation}
\gamma={1\over R'C'}.
\end{equation}
In this rescaled form, our model is identical to one employed recently in nanomechanics by Babourina-Brooks, Doherty, and Milburn\cite{BBDM}.  In fact, its essential features were already calculated thirty years ago by Drummond and Walls\cite{DW}.  In this paper, we will explore this rich solution and provide physical insights into some of its many interesting regimes.

\subsection{spectroscopy} 

We will model the ``spectroscopy'' experiment as done in
\cite{SQUIDless} and in many other recent studies of superconducting
qubit systems.  Spectroscopy is measured as the magnitude of the ac
output voltage signal at the same frequency as the input.  For the
model above, this will be proportional to the expectation value of $a$
in steady state:
\begin{equation}
\langle a\rangle={\rm Tr}(a\rho_0)
\end{equation}
Where $\rho_0$ is the steady state response:
\begin{equation}
{d\rho_0\over dt}=0
\end{equation}
We can write the Lindblad master equation in the form
\begin{equation}
{d\rho\over dt}=S[\rho]
\end{equation}
where $S$ is a linear superoperator, generating a TCP (trace preserving completely positive) map $e^{St}$.  The resulting equation for the steady state,
\begin{equation}
S[\rho_0]=0
\end{equation}
states that $\rho_0$ is obtained as the zero eigenvector of the
linear non-Hermitian (super)operator $S$.  It is this point of view
that is employed in the numerical studies that we present below.

But numerics are only necessary for developing physical explanations,
since the spectroscopy calculation itself, remarkably, has a closed
form solution.  Quoting Eq. (5.17) of \cite{DW}:
\begin{equation}
\langle a\rangle ={\epsilon\over\Delta-i\gamma/2}{\,_0F_2\left( {\Delta+\chi-i\gamma/2\over\chi},{\Delta+i\gamma/2\over\chi},{2\epsilon^2\over\chi^2}\right)\over \,_0F_2\left({\Delta-i\gamma/2\over\chi},{\Delta+i\gamma/2\over\chi},{2\epsilon^2\over\chi^2}\right)}\ .
\label{closed}
\end{equation}
Here $\,_0F_2$ is a standard hypergeometric function. 

\section{Results}

\begin{figure}[htp]
	\centering
	\subfloat[$\gamma=2$]{\includegraphics[height=4.5cm]{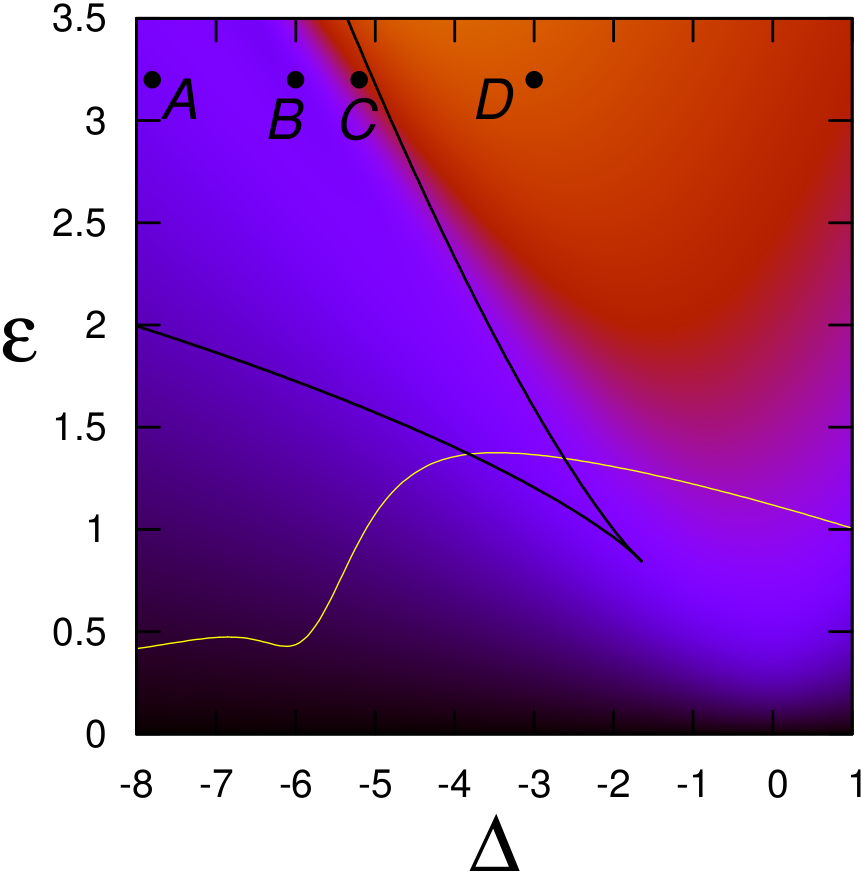}}
\ \ \ 	\subfloat[$\gamma=0.3$]{\includegraphics[height=4.5cm]{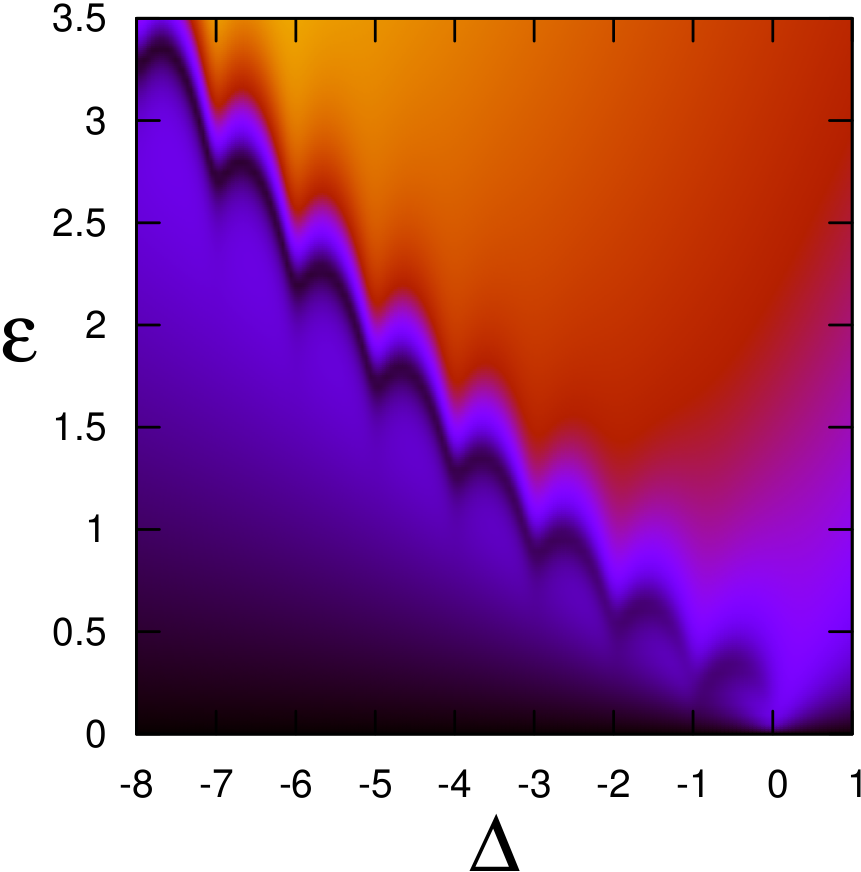}}
\ \ \ 	\subfloat[$\gamma=0.01$]{\includegraphics[height=4.5cm]{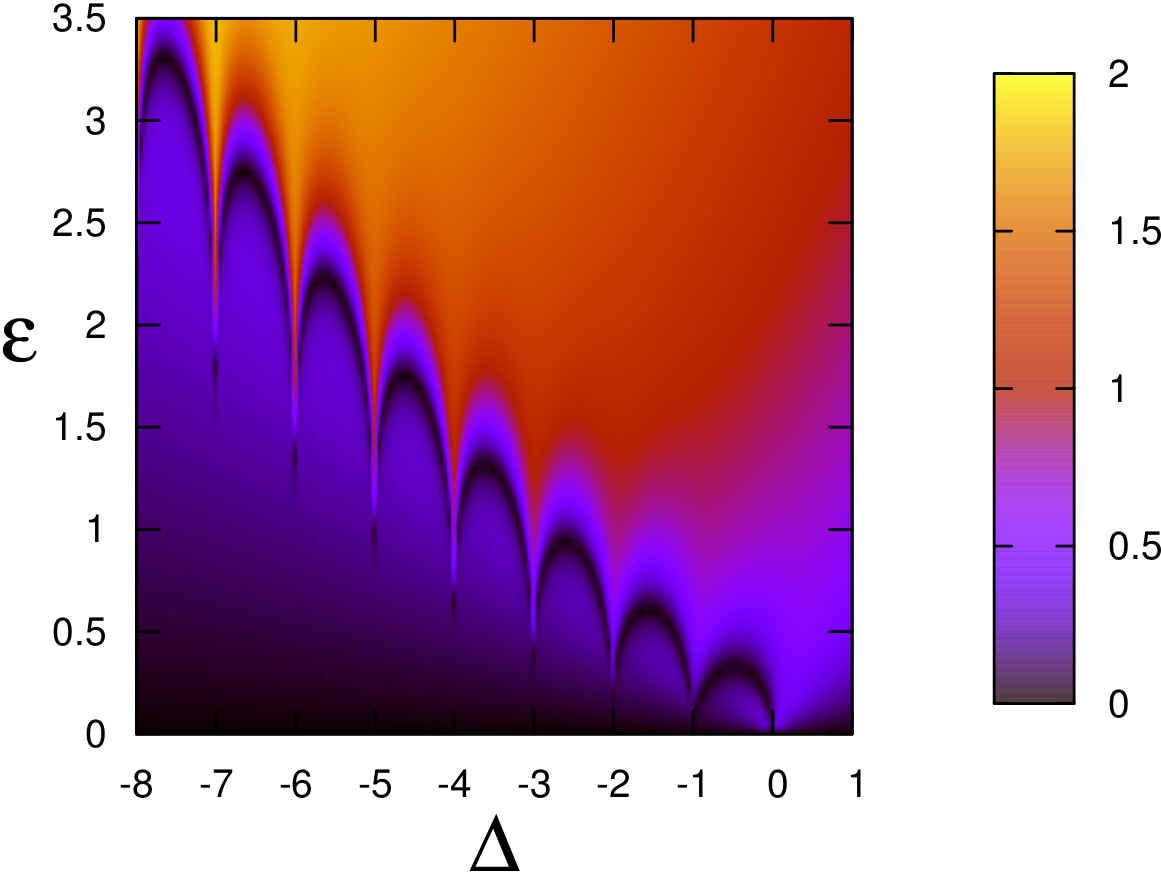}}\\
        \subfloat[$\gamma=0.01$]{\includegraphics[height=5cm]{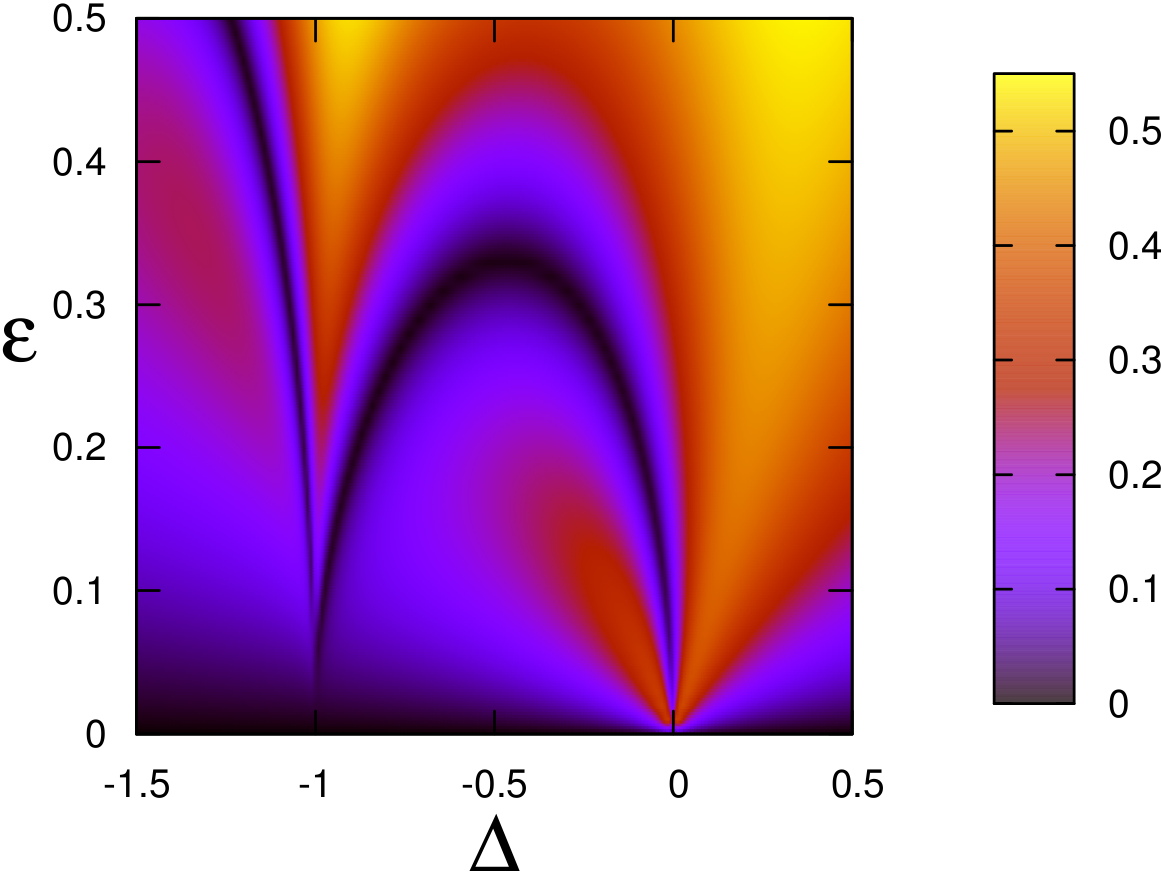}}
	\caption{Response amplitude $|\langle a\rangle |$ as a function of detuning $\Delta$ and excitation amplitude $\epsilon$, all for anharmonicity parameter $\chi=1$.  (a) Moderately high damping ($\gamma=2$).  Red line: scan across the response at $\epsilon=3.2$.  Points A, B, C, and D along this scan path are examined further in Fig. \ref{fig:wignersurvey}.  The trough at point $B$ is easily visible in the line scan.  Dark lines show the boundary of the metastable points \protect\cite{bifur} for the classical Duffing oscillation with these same parameters. (b) Response for moderate damping $\gamma=0.3$.  Trough has become more prominent but is now scalloped, indicating the development of discrete quantum resonances.  (c) Response for low damping $\gamma=0.01$.  Quantum resonant structure becomes very prominent.  (d) A closer view of the first resonant structure in the low damping case $\gamma=0.01$.}
	\label{fig:response}
\end{figure}

For three different damping parameters $\gamma$, Fig. \ref{fig:response} shows $|\langle a\rangle |$, which is proportional to the magnitude of the transmitted voltage in a spectroscopy experiment.  The plots are shown as a function of $\Delta$, the detuning away from the harmonic resonance frequency, and $\epsilon$, the drive intensity.  These quantities are normalized to the anharmonicity parameter $\chi$ ({\em i.e.}, $\chi=1$ in Eq. (\ref{RWAeq})).  The three panels show three different settings of the damping parameter (again, relative to $\chi$).  For the largest $\gamma$ shown ($\gamma=2$ in part (a)) the response is in some sense classical; no structure involving transitions between individual quantum mechanical levels is present at any excitation level.  As we will discuss, though, there are many aspects of this result that are not classical in any naive sense.  

The principal feature of Fig. \ref{fig:response}(a) at moderate excitation levels ($\epsilon\gtrsim 1$) is a fairly abrupt upward step in the response, with a brief dip in between.  This dip or trough has been noted previously\cite{DW,TroughExpt,TroughTheory}, and is prominently seen in recent experiments; we will discuss its origin shortly.  As $\gamma$ is decreased (Fig. \ref{fig:response}(b)), this trough develops ripples that are synchronized with the quantum transitions in the anharmonic ladder of states.  At very small $\gamma$ (Fig. \ref{fig:response}(c)) these ripples resolve themselves into spectroscopic resonant peaks, which merge into the ``high" continuum at large $\epsilon$.  As we will discuss later, it is only in this third setting that it is meaningful to say that the anharmonic resonator behaves as a qubit, for the most part only at small excitation levels $\epsilon\lesssim\chi$ and near zero detuning $|\Delta|\lesssim\chi$.

\section{Results: high damping regime}

The true classical behavior would be quite different from what is seen
in Fig. \ref{fig:response}(a), as was noted already in \cite{DW}.
Inside the two lines shown in Fig. \ref{fig:response}(b), these are
two classical steady state solutions -- the dynamics bifurcates at the
meeting of these two lines, and is hysteretic above the bifurcation.
A quantum treatment, even one with large loss, cannot show this
bifurcation, although the quantum solution is in fact very close to
the classical solution in the ``low" region beneath the lower
bifurcation line, and in the ``high" region above the upper
bifurcation\cite{DW}.  Qualitatively, this difference can be ascribed
to the possibility of transitions between the two classical steady
states; mathematically, it arises because the superoperator $S[\rho]$
has a unique zero eigenvalue.  Classical hysteretic behavior can be
and is recovered in a quantum treatment of the transient behavior, as
studied recently in \cite{metastab}.

Collapse of hysteresis due to quantum effects is actually well known
\cite{FreedmanSarachik}.  Special to the anharmonic resonator system,
however, is the robust trough easily seen in
Fig. \ref{fig:response}(a).  As originally noted in \cite{DW}, the
trough can be reasonably explained by the observation that the
classical ``high" and ``low" states have almost opposite phases, since
the first is essentially a resonant response, while the second is
non-resonant.

\begin{figure}[htp]
\begin{flushleft}
	\subfloat[Point $A$, $\Delta=-7.8$]{\includegraphics[height=6cm]{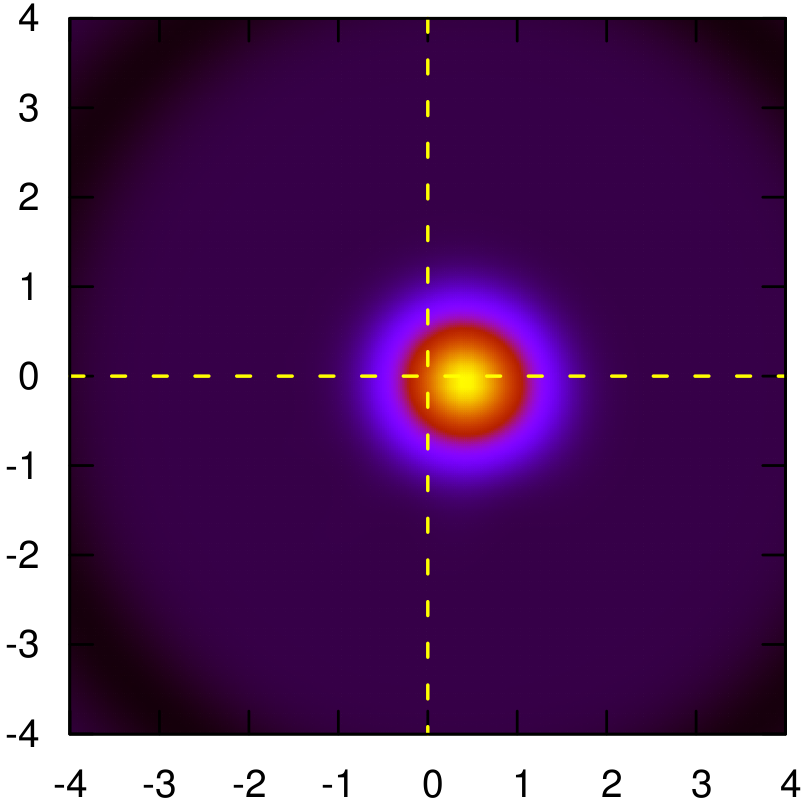}}
	\subfloat[Point $B$, $\Delta=-6.0$]{\includegraphics[height=6cm]{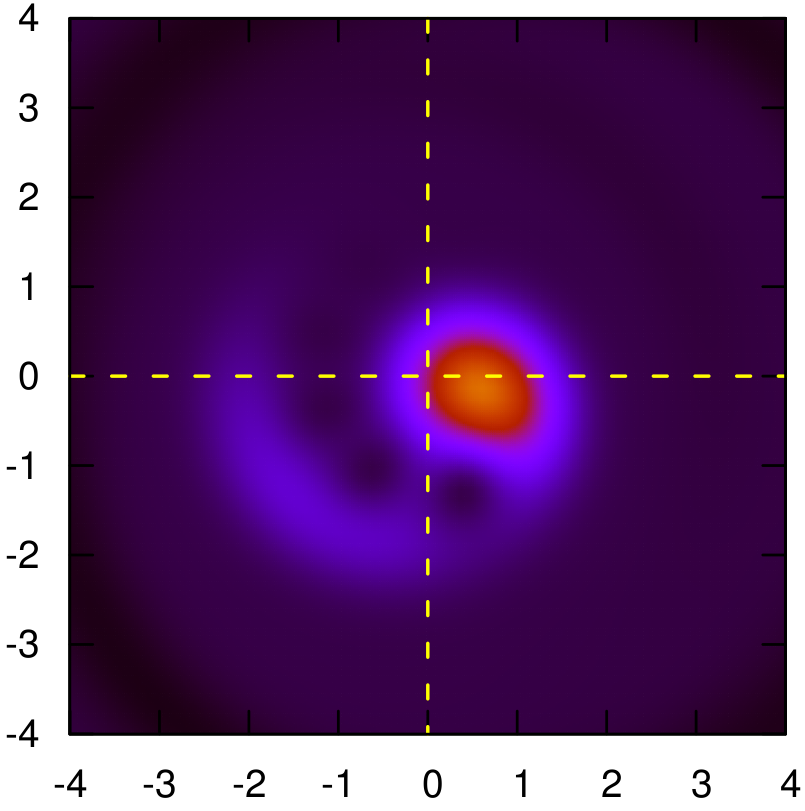}}\\
	\subfloat[Point $C$, $\Delta=-5.2$]{\includegraphics[height=6cm]{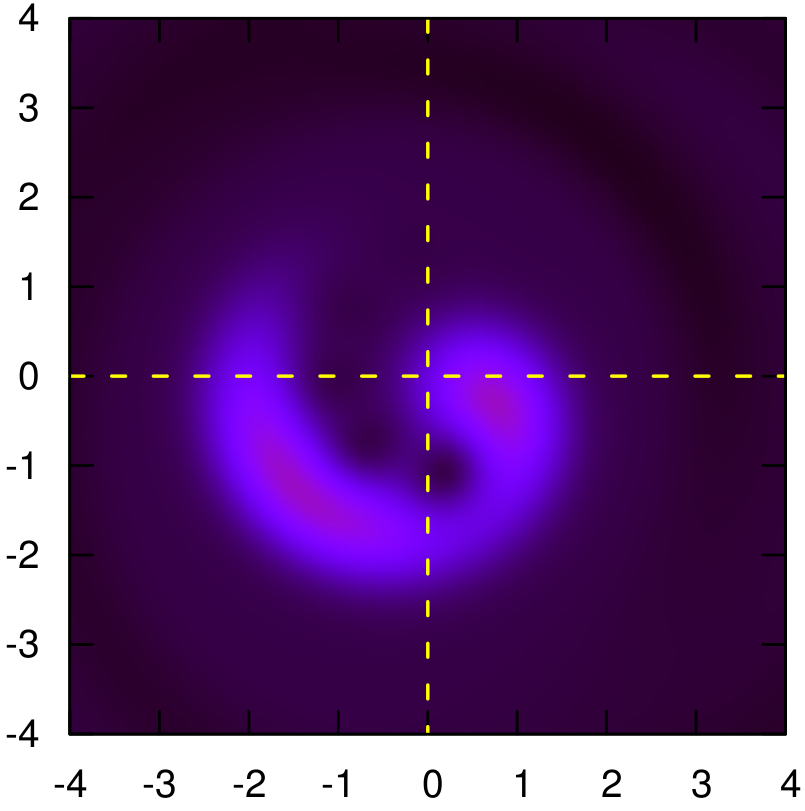}}
	\subfloat[Point $D$$, \Delta=-3.0$]{\includegraphics[height=6cm]{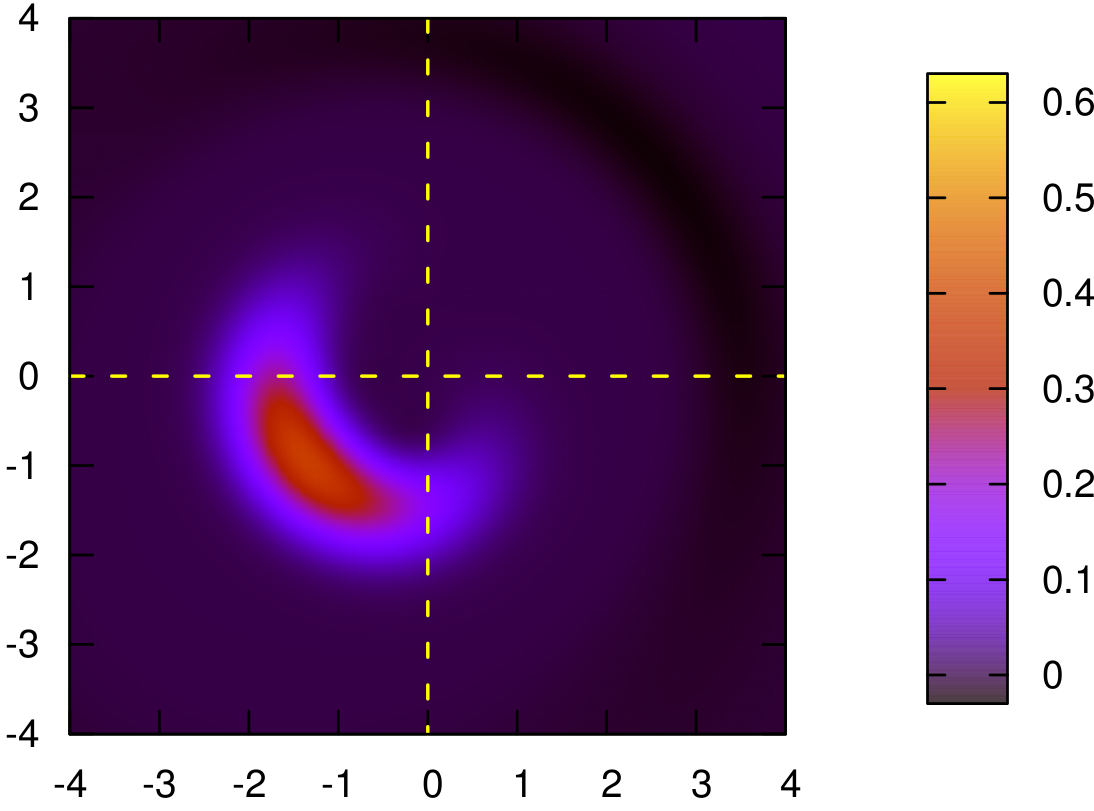}}
\end{flushleft}
\caption{Wigner functions for four points shown in Fig. \ref{fig:response}. }
	\label{fig:wignersurvey}
\end{figure}

In the simplest view of the trough, the state intermittently switches
back and forth between the high and low Gaussian state.  While this
gives a rough description of the situation at point $B$ in
Fig. \ref{fig:response}(a), the state there is in fact not purely a
mixture of a low-amplitude with a high-amplitude Gaussian state.  We
explore this transition by examining the Wigner function $W$ of the
steady state $\rho_0$ 
as we pass through the ``trough" by varying the detuning $\Delta$.  We
use the standard formula\cite{Wigner}
\begin{eqnarray}
W(\alpha)&=&{\rm Tr}\left( D(-\alpha)\rho_0D(\alpha)\Pi\right)\nonumber\\  
D(\alpha)&=&\exp\left(\alpha a^\dagger-\alpha^* a\right),\,\,\,\Pi=\sum_{n=0}^\infty (-1)^n\ket{n}\bra{n}
\end{eqnarray}
with complex displacement $\alpha$.  The simplest is the Wigner
function at the ``low response" point $A$
(Fig. \ref{fig:wignersurvey}(a)).  To the eye, this state is nearly a
coherent state slightly displaced ({\em i.e.}, with substantially less than
one photon of excitation).  In fact, the state has some excess noise
({\em i.e.}, has a somewhat broader Wigner function than a coherent state),
which can be measured by the von Neumann entropy $S_{\rm
  vN}(\rho)=-{\rm Tr}\rho\log\rho$.  At point $A$, $S_{\rm
  vN}(\rho_0)=0.03$ bits.

The Wigner function at the ``high response'' point $D$ departs further
from the simple classical view.  This Wigner function
(Fig. \ref{fig:wignersurvey}(d)) of course has a larger
$|\langle\alpha\rangle |$ than in point $A$, but it is much farther
from a coherent state.  Despite its appearance, it is not number
squeezed (the radial width is not smaller than a coherent state).  It
is perhaps better described as a coherent state with significant
excess phase noise.  The entropy of this state is $S_{\rm vN}=0.85$
bits.

The Wigner function in the transition region deserves considerable
discussion. We will focus in detail on point $C$. The classical view
that the state is a mixture of two classical responses is not entirely
incorrect.  But it fails to capture many of the aspects of the actual
response. The system clearly spends appreciable time at phases
intermediate between the high and low response, with $\bar{n}=2.56$, so
it should not be viewed as quickly switching between the two.  The
state has high entropy, as a classical mixed state would, but its
precise value ($S_{\rm vN}=1.74$) gives us more
information about the nature of the state, as we explain now.

\begin{figure}[htp]
	\centering
	\subfloat[$\rho_-$]{\includegraphics[height=6cm]{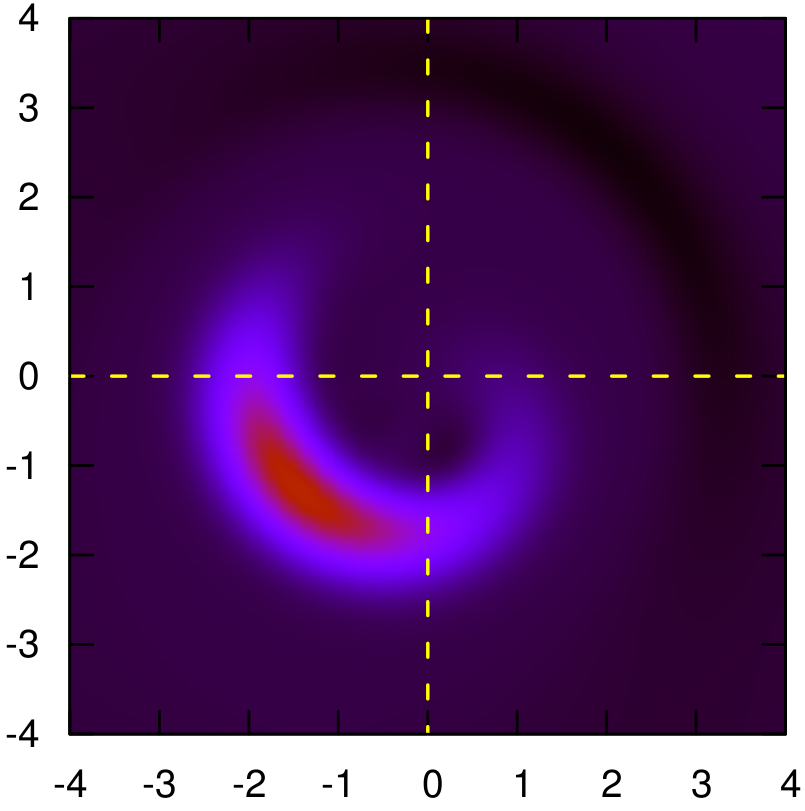}}
	\subfloat[$\rho_+$]{\includegraphics[height=6cm]{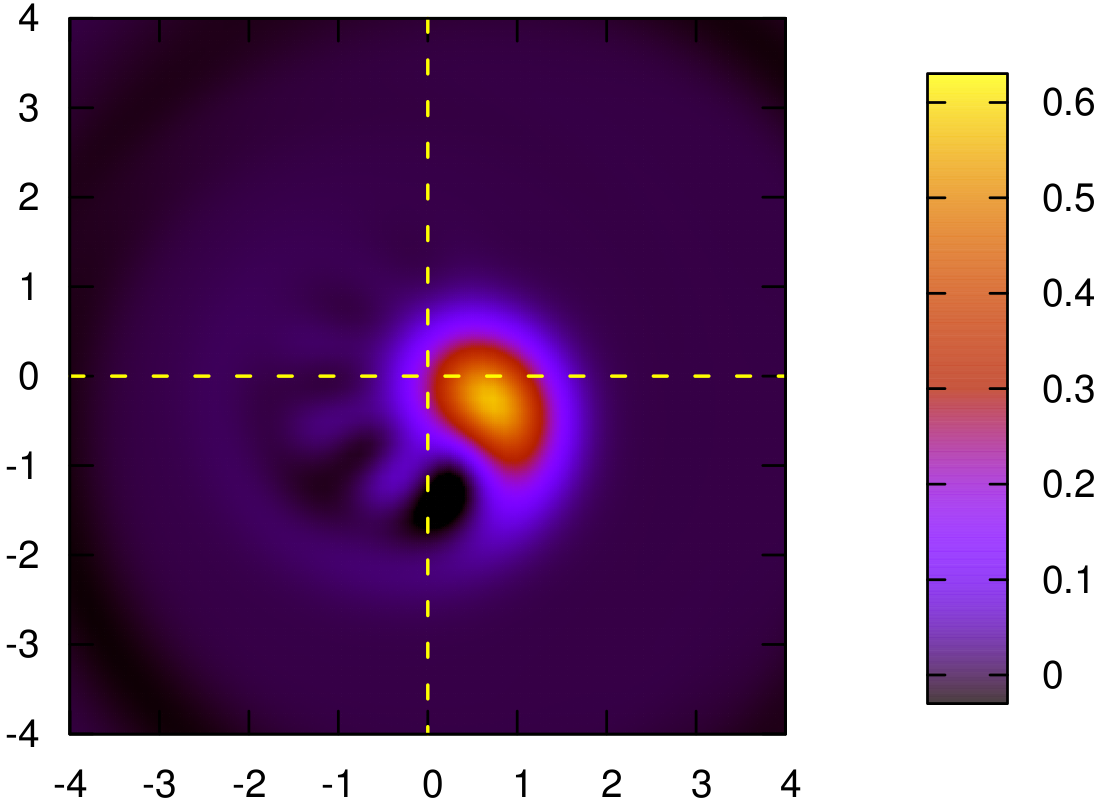}}
	\caption{Wigner functions of the extremal metastable responses at point $C$ ($\Delta=-5.2$)  in the response diagram, showing that the picture that the Wigner function of the response (Fig.~\ref{fig:wignersurvey}(c)) involves two possible metastable states is approximately, but not precisely, correct.  Color scale is the same as in Fig. \ref{fig:wignersurvey}.}
	\label{fig:rhoplusminus}
\end{figure} 

\begin{figure}[htp]
	\centering
		\includegraphics[width=14cm]{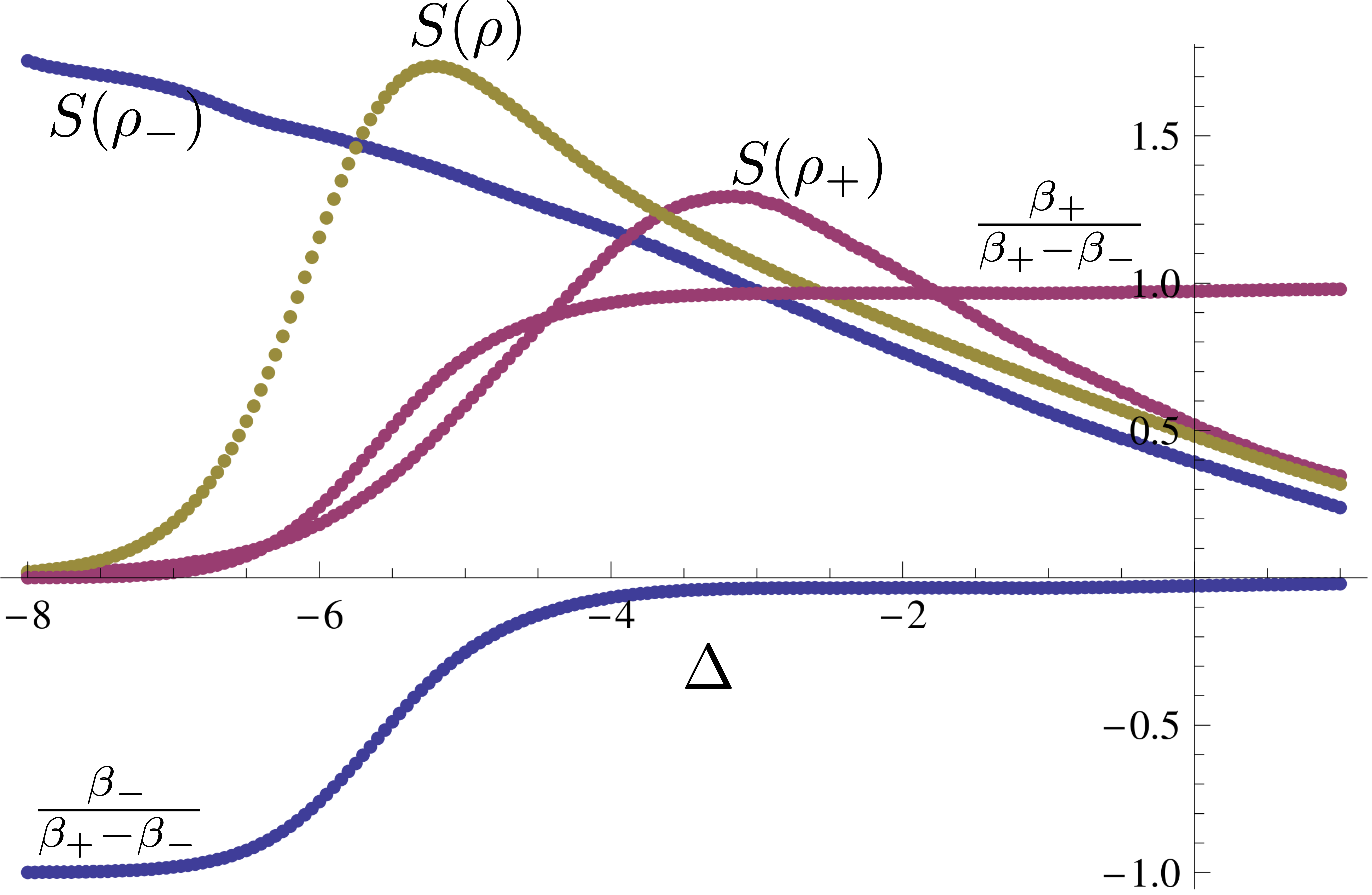}
                \caption{von Neumann entropies for the steady state
                  $\rho$ and the extremal metastable states $\rho_+$
                  and $\rho_-$, as a function of $\Delta$ for
                  $\epsilon=3.2$.  Also shown are the coefficients
                  $\beta_\pm$, suitably normalized, in
                  Eq. (\ref{extremes}). }
	\label{fig:entropyplus}
\end{figure}

The mixing picture is illuminated by a computation which
pulls out a version of the two ``classical response'' states from the
full state of Fig. \ref{fig:wignersurvey}(c); this is shown in
Fig. \ref{fig:rhoplusminus}.  A vestige of the bistability of the
classical response is clearly reflected in the eigenstructure of the
superoperator $S$ at the operating point $C$.  The second smallest
eigenvalue if $S$ is at $\lambda_1=-0.215$, while the third eigenvalue
is at $\lambda_2=-2.204$.  It is the presence of two very nearby low
lying eigenvalues $\lambda_{0,1}$, with a large separation from the
rest of the spectrum, is a reflection of the classical bistability.
It is straightforward to calculate this second eigenvector $|1)$
a.k.a. $\delta\rho_1$: $S|1)=\lambda_1|1)$.  We use the notation
$\delta\rho$ as a reminder that while this eigen-matrix is Hermitian
(because it is associated with a real eigenvalue, rather than with a
complex-conjugate pair\cite{supopI}), it is traceless.  But the
important object we can construct from $\delta\rho_1$ is
\begin{equation}
\rho_0+\beta\delta\rho_1.\label{range}
\end{equation}
For real $\beta$.  There will be a range of $\beta$s, including
$\beta=0$, for which the operator will be positive semidefinite.  We
find that the two extremal operators, the one with the smallest
possible $\beta=\beta_-<0$ and the largest possible $\beta=\beta_+$>0,
are the ones corresponding to the two classical solutions.
Fig. \ref{fig:rhoplusminus} shows the Wigner functions of these two
density operators,
\begin{equation}
\rho_\pm=\rho_0+\beta_\pm\delta\rho_1\label{extremes}
\end{equation}
We can see that this does a reasonably good job of separating the
response into a high-amplitude and a low-amplitude part.  This is also
confirmed by the calculation of the von Neumann entropy of the range
of states Eq. (\ref{range}) (Fig. \ref{fig:entropyplus}).  The two
extremal states $\rho_\pm$ are those of the lowest entropy.
Classically, the states in the interior of this range (including
$\rho_0$) should have an excess entropy above the weighted average of
the extremal states (see Fig. \ref{fig:entropyplus}) equal to the
ordinary binary entropy of mixing $H(x)\equiv-x\log x-(1-x)\log(1-x)$.
In fact Fig.  \ref{fig:entropyplus} differs from this classical
situation in two respects: 1) The maximal entropy of mixing is not 1
bit, but about 0.74 bits.  2) The excess entropy is nearly
proportional to $H(x)$, but not exactly so, reflecting a quantum
effect, the noncommutivity of the extremal states $[\rho_-,\rho_+]\neq
0$.

\section{Results: Low damping regime}

We now turn our attention to the lightly damped regime (by which we will
mean $\gamma\lesssim\chi$).  We expect that it is in this regime that
our anharmonic oscillator ``is" or ``behaves like" a qubit.  We will
explore the light that our response calculation can shed on the
question of when such an oscillator is or is not a qubit.

The outstanding qualitative feature of the response as $\gamma$ is
lowered (as in the series in Fig. \ref{fig:response}) is that
stalactites form, descend, and sharpen from the bottom of the
high-response continuum.  The continuum itself, at least in the
vicinity of the classical upper bifurcation line of
Fig. \ref{fig:response}(a), is not much affected by the lowering of
$\gamma$.  The trough remains an important feature, becoming first
slightly, and then strongly, scalloped, finally breaking up into
segments between one stalactite and the next.

Given that the quantum energy levels $\omega_n$ for the quartic
oscillator obey to good approximation the simple rule
\begin{equation}
\omega_{n+1}-\omega_n=\omega_0+n\chi,
\end{equation}
it is natural to associate the resonant structure at zero detuning
with the ``qubit'' $0\rightarrow 1$ transition, the structure at
detuning $\Delta=-\chi$ with the $1\rightarrow 2$ transition, that at
$\Delta=-2\chi$ with the $2\rightarrow 3$ transition, and so forth.
We might be tempted to guess that excitation detuning $\Delta=-n\chi$,
the system is acting like a qubit involving the levels $\ket{n}$ and
$\ket{n+1}$.  As we will seen this is most definitely not the case,
except for $n=0$.  A detailed inspection of the $n=0$ ``stalactite''
shows that its structure is quite distinct from that of all the
others.  In fact, the excitation structures at higher $n$ are
indicative of interesting multi-level behavior, as we will see.

\subsection{A perturbation theory}

We will do a perturbation analysis in $\epsilon$ to elucidate the
nature of the resonant structures that appear as the excitation level
is increased.  While much can be learned simply by Taylor expansion of
the exact expression for $|\langle a\rangle |$, much more is revealed
by developing perturbative expressions for the eigenoperator of the
superoperator $S$.

To develop this, we first record an explicit matrix form for this superoperator:
\begin{eqnarray}
S_{kl,ij}\rho_{ij}&=&\{-i[\Delta(k-l)+\chi[k(k-1)-l(l-1)]]\delta_{ki}\delta_{lj}\nonumber\\
&&-i \epsilon(\sqrt{k+1}\delta_{k,i-1}\delta_{lj}+\sqrt{k}\delta_{k,i+1}\delta_{lj}-
\sqrt{l}\delta_{ki}\delta_{j+1,l}-\sqrt{l+1}\delta_{ki}\delta_{j-1,l})\nonumber\\
&&+{\gamma\over 2}[2\sqrt{(k+1)(l+1)}\delta_{k+1,i}\delta_{l+1,j}-(k+l)\delta_{ki}\delta_{lj}]\}\rho_{ij}
\end{eqnarray}
S is sparse, but of course it is not, as a matrix, Hermitian. We
analyze the steady state response by a perturbation approach.  we can
separate $S$ thus:
\begin{equation}
S=S_0+\epsilon V
\end{equation}
$S_0$ can be analytically diagonalized -- it has a ``bidiagonal'' form
as a matrix.  It is convenient to index its eigensolutions by two
integers $n$ and $q$.  Since $S_0$ is non-Hermitian, it is necessary
to distinguish its right from its left eigenvectors,
\begin{equation}
S_0|nq_R)=\lambda_{nq}|nq_R)
\end{equation}
\begin{equation}
(nq_L|S_0=(nq_L|\lambda_{nq}
\end{equation}
Recall that the left eigenvectors are not the adjoint of the right
eigenvectors, but the matrix formed by the left eigenvectors is the
inverse of the matrix of right eigenvectors.  The physical steady
state response to zeroth order in $\epsilon$ is the eigenvector
$|00_R)$.

$S_0$ has both real and complex eigenvalues, for both of which ${\rm
  Re}\lambda_{nq}<0$ for $\epsilon\neq 0$ (except for $\lambda_{00}$
which is always equal to zero).  It is straightforward to confirm the
following: All the complex eigenvalues of $S_0$ may be written
\begin{equation}
\lambda_{nq}=-\left(q+{n\over 2}\right)\gamma-ni\Delta-n(n-1+2q)i\chi
\end{equation}
with integers $q\ge 0$, $n>0$.  The right eigenvectors $|nq)$ (a.k.a. $\rho_{nq}$, see Ref. \cite{Caves}) are operators on the Hilbert space with matrix elements
\begin{equation}
\langle s|\rho_{nq}|t\rangle=\delta_{s,t+n}{\left(-1-2n{i\chi\over\gamma}\right)^t
\over(q-t)!\sqrt{s!t!}}\ .
\end{equation}
Each has a complex conjugate partner $\lambda_{{\bar n}q}=\lambda^*_{nq}$, $\rho_{{\bar n}q}=\rho^\dagger_{nq}$.  The real eigenvalues and right eigenvectors are given by the same formulas with $n=0$.  We do not record the left eigenvectors $(nq|$ here, but they are as straightforward to write down as the right eigenvectors. 

The right eigenvector $|00_R^{\rm full})$ of the full superoperator can be conveniently developed using Brillouin-Wigner perturbation theory\cite{Baym}.  Going over to a schematic single-integer indexing of the eigenstates:
\begin{equation}
|0_R^{\rm full})=|0_R)+\epsilon \sum_{k>0}{|k_R)(k_L|V|0_R)\over E_0-\lambda_k}+\epsilon^2 \sum_{k>0}\sum_{j>0}{|k_R)(k_L|V|j_R)(j_L|V|0_R)\over
(E_0-\lambda_k)(E_0-\lambda_j)}+...
\end{equation}
The Brillouin-Wigner formula is frequently not used because the exact eigenvalue $E_0$ is unknown.  But in this application, of course, $E_0=0$ exactly; so this series is actually very useful, and its first few terms are readily calculated.

\subsection{Characteristics of the resonant response}

\begin{figure}[htp]
	\centering
		\includegraphics[height=6cm]{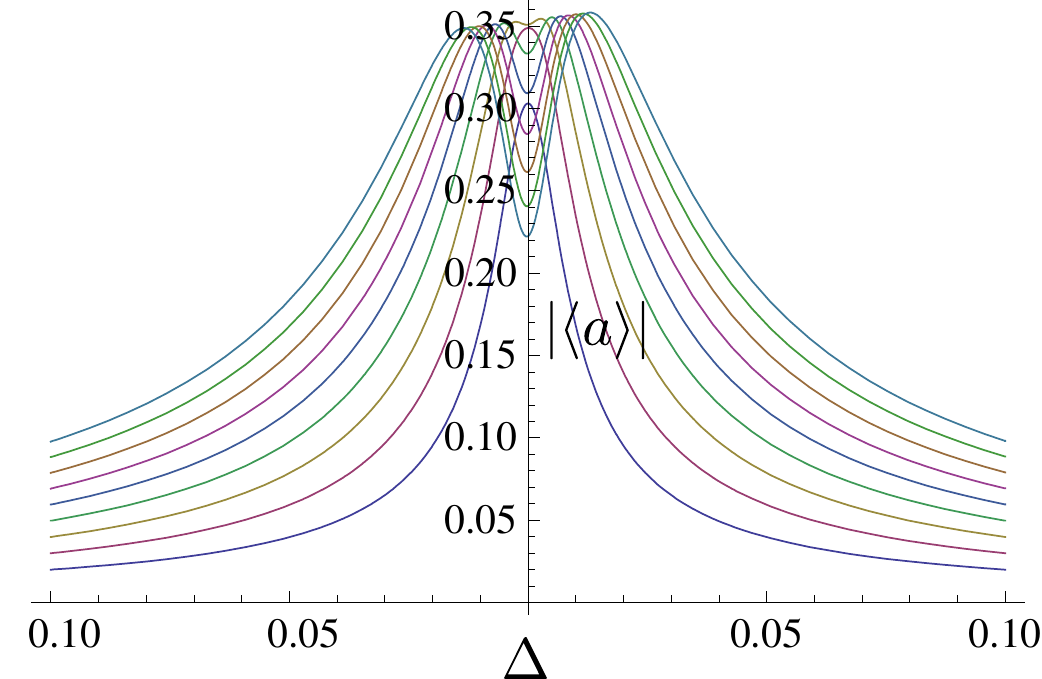}
                \caption{Response $|\langle a\rangle |$ for low
                  damping, showing the emergence of the line at zero
                  detuning. Here $\gamma=.01$, $\chi=1$, and from
                  bottom to top,
                  $\epsilon=.002,.003,.004,.005,.006,.007,.008,.009,.01$.
                  At the low excitation level the response is
                  Lorentzian and corresponds exactly to the linear
                  response of a qubit; the higher lying levels play no
                  role in the response.  The hole in the center of the
                  line at the highest $\epsilon$ is also a qubit
                  effect, showing the incipient saturation of the
                  qubit response.  The asymmetry of the line around
                  zero detuning occurs because of participation of
                  higher levels, and would be the first indication
                  that the system is not a qubit as the excitation
                  level is raised.}
\label{zerdet}
\end{figure}

\begin{figure}[htp]
	\centering
        \includegraphics[height=6cm]{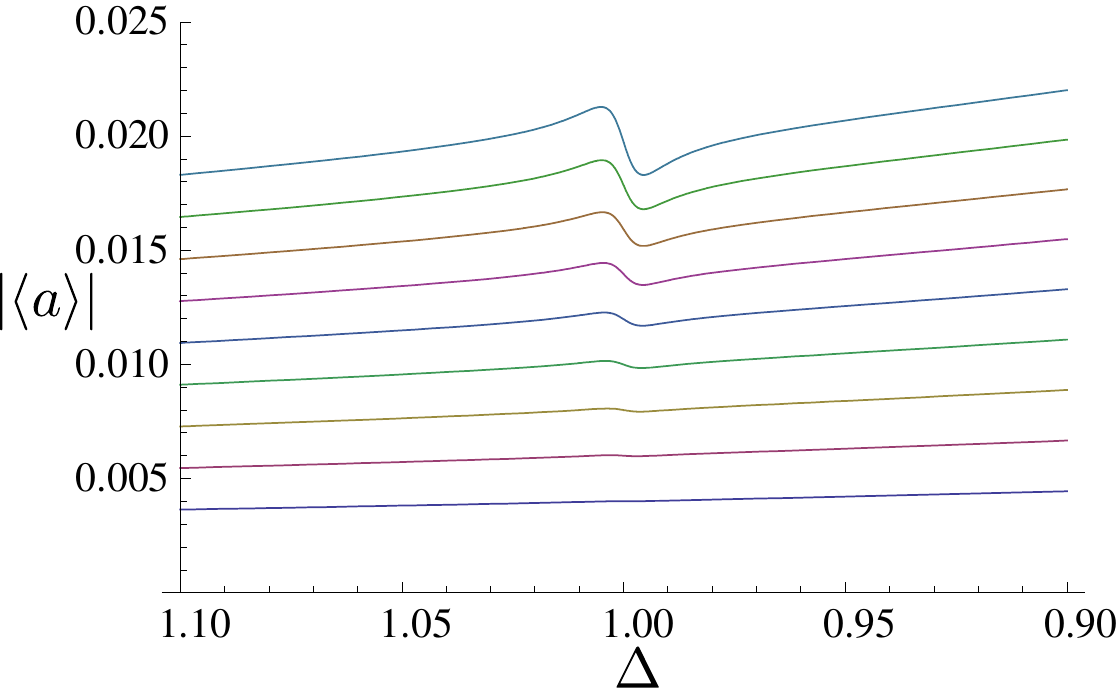}
	\caption{Response $|\langle a\rangle |$ for low damping,
          showing the emergence of the line at detuning
          $\Delta=-\chi$.  Here $\gamma=.01$, $\chi=1$, and, from
          bottom to top,
          $\epsilon=.004,.006,.008,.010,.012,.014,.016,.018,.02$.  The
          structure emerges as a Fano resonance out of the continuum
          of the far off resonant response at zero
          detuning.}\label{onedet}
\label{zerdet2}
\end{figure}

Using this perturbation theory, we can write out in matrix representation (in the number basis) the steady state response in orders of $\epsilon$:
\begin{equation}
\rho_0^{\rm approx}\approx\left(\begin{array}{cccc}1-O(\epsilon^2)&\cdot&\cdot&\cdot\\
\begin{array}{l}{2\epsilon\over-2\Delta+i\gamma}\,\,+\\
{8\epsilon^3(4\chi+2\Delta-i\gamma)\over
(2\Delta-i\gamma)^2(2\chi+2\Delta-i\gamma)(2\Delta+i\gamma)}\end{array}&
O(\epsilon^2)&\cdot&\cdot\\O(\epsilon^2)&{-8\epsilon^3\over
(2\Delta-i\gamma)(2\chi+2\Delta-i\gamma)(2\Delta+i\gamma)}&O(\epsilon^4)&\cdot\\
O(\epsilon^3)&O(\epsilon^4)&O(\epsilon^5)&O(\epsilon^6)\end{array}\right)\ .
\label{bigmatrix}
\end{equation}
The matrix is Hermitian, and we have omitted the entries above the
diagonal.  We have only shown in detail the matrix elements
$\bra{1}\rho_0^{\rm approx}\ket{0}$ and $\bra{2}\rho_0^{\rm
  approx}\ket{1}$, since these are the only ones that contribute to
the the observable $\langle a\rangle$; but, as we note, there are many
non-zero unobservable components to this steady-state density
operator.

We also note the expression for the total response that can be
extracted either from this or via a Taylor expansion of
Eq. (\ref{closed}):
\begin{equation}
\langle a\rangle={2\epsilon\over 2\Delta-i\gamma}-{32\chi\epsilon^3\over
(2\Delta-i\gamma)^2(2\chi+2\Delta-i\gamma)(2\Delta+i\gamma)}+O(\epsilon^5) \ .
\label{expanding}
\end{equation}

First we turn our attention to the resonant structure near zero
detuning $\Delta=0$, shown in Fig.~\ref{zerdet}.  For small
$\epsilon$, as the expression in Eq. (\ref{expanding}) shows, the
response is a simple Lorentzian, just as expected for a driven
two-level system.  In this regime, the oscillator behaves exactly like
a qubit, which would, in the transient regime, respond as any two
level system would to Rabi excitation or to Ramsey or spin-echo
protocols.  When $\epsilon$ become of order $\gamma$, new structure
appears.  The top of the Lorentzian peak becomes flat and then exhibits
a growing hole.  One might say that at this point the higher levels of
the oscillator are participating, but in fact this is very close to
the behavior shown by a two level system in the regime of spectroscopy
probed at high power.  As recently shown in \cite{BishChow},
saturation of the two level system causes just this same evolution
into a double peaked structure.  But some features of this regime are
indicative of the participation of higher levels.  The development of
line asymmetry is definitely ascribed to higher levels; this asymmetry
diminishes in the limit of large $\chi$.

Mathematically, we see from Eq. (\ref{bigmatrix}) that these new
structures in the zero-detuning line all arise from the
$O(\epsilon^3)$ contributions.  Note that only odd orders contribute
observably to $\langle a\rangle$, consistent with the fact that the
response should be an odd function of $\epsilon$.  The $O(\epsilon^3)$
response involves all of the first four levels, so in some sense the
line at this excitation level is indicative of four-level system
response.  But it is not clear that one need adopt this point of view,
since the 0-3 matrix element of the steady state response is not
visible to our problem, although one could construct a more complex
observable that is sensitive to it.

We next turn to an examination of the resonant structure near
$\Delta=-\chi$, at the ``$1\rightarrow 2$ transition''.  The linear
response has no resonant structure at this detuning, but the
$\epsilon^3$ part of the response is resonant here.  But the linear
response has a uniform far off resonant background at this detuning,
and resonance emerges our of this background (see Fig. \ref{zerdet2}).
Thus, the behavior here is very different in detail from the
no-detuning line, although mathematically, they arise from two
different poles of the same matrix elements in Eq. (\ref{bigmatrix}).

Furthermore, it is easy to show that because of interference of the
resonant response with the background, the line that emerges in
Fig. \ref{onedet} has a Fano line shape\cite{Fanoline}
\begin{equation}
|\langle a\rangle |\sim|\langle a\rangle |_{background}+C{(x-q)^2\over x^2+1}
\end{equation}
with suitably normalized frequency $x$.  The Fano parameter $q$ is easy to calculate:
\begin{equation}
q={-2\chi\over\sqrt{2\chi^2+\gamma^2}-\gamma}
\end{equation}
Since we are in a regime where $\gamma<<\chi$, this Fano parameter is
close to $-1$.  $q=-1$ is among the most asymmetrical of the Fano line
shapes \cite{Fano61}, in striking contrast to the initial symmetry of
the distorted zero-detuning line.  This Fano line exhibits a trough on
the high-frequency side of the line.  This is the beginning of the
trough that forms a scallop from one ``stalactite'' to the next, and
so it is the structure that has continuously evolved from the
semiclassical trough that occurs at high damping.  It is curious that
the classical interference arising from intermittency at high damping
evolves continuously into a Fano destructive interference, which we
think of as entirely quantum mechanical.

We will say much less about the resonant structures that arise at
higher detuning, except to make a couple of observations: All higher
lines emerge in a similar way to the $\Delta=-\chi$ line, with a Fano
line shape with the same asymmetry (so, evidently, a similar $q$,
which we have not calculated).  Each emerges first at successively
higher order in the response; so, the resonant line at $\Delta=-n\chi$
arises first at order $O(\epsilon^{2n+1})$.  The calculation shows
that the first appearance of the line from the background (as measured
by the first appearance of a zero slope near the resonance frequency)
occurs at an excitation level that goes like
\begin{equation}
\epsilon_{onset}=f(n)\chi^{1-{1\over n}}\gamma^{1\over n},\label{scaler}
\end{equation}
with some increasing function $f(n)$. We note that the dominant
background is always the off resonant response of the zero-detuning
line, rather than the background of any nearer by lines, no matter how
large the value of $n$.  So, from Eq. (\ref{scaler}) we see that it is
only the $n=0$ and $n=1$ lines that first show nonlinear structure at
$\epsilon\sim\gamma$; structures at all higher detunings occur only at
higher power ({\em e.g.}, for $n=2$ the onset is at
$\sqrt{\chi\gamma}$).

\section{Discussion}

It is clear that the present study is only a piece of the full picture
in the theoretical modeling of the complete input-output relations for
the excitation of qubit structures.  The potential complexity of the
decohering environment of these structures has barely been touched on
here, and remains a frontier area of research. For actual
applications, more of the essential further work will be in the time
domain rather than the frequency domain; pulsed excitation of qubits,
both for gate operations and for measurements, will definitely be the
study of more extensive modeling as this experimental art develops
further.

Still, the present simple study has revealed an intricate complex of
phenomena.  One example of a further study that it possible as a
direct follow on to this work is the more detailed examination of the
validity and limits of the rotating wave approximation.  The
simplicity of the mathematics here should permit a definite if limited
answer to questions about this.  These questions often do not have
such a clear answer in a more general setting.

Another question raised by the present study concerns the emergence of
classical bistability.  As we have seen, the quantum theory says that
there is a continuum of states (convex combinations of $\rho_+$ and
$\rho_-$) that are metastable.  Which two of them does the classical
physics pick out?  We have offered a conjecture here that the ones of
locally minimum von Neumann entropy (which happen to be $\rho_+$ and
$\rho_-$ themselves) are the classical states.  Their low entropy is
associated (albeit indirectly) with their high degree of localization
of the Wigner function
in the complex $\alpha$ plane.  This in turn could make them
candidates as ``pointer states.''  That is, it is possible that the
right view of the quantum environment would make these states emerge
naturally.  Thus, even the most fundamental questions can have a
profitable launching point from our simple model calculations.

{\bf Acknowledgments:} The authors would like to thank Chad Rigetti for
helpful discussions and comments on the text.  JAS has been partially supported
by the IARPA MQCO program under contract no. W911NF-10-1-0324.

\end{document}